\documentclass[a4paper, 10 pt, conference]{ieeeconf}
\IEEEoverridecommandlockouts
\usepackage{amsfonts,amsmath,amssymb} 

\newcommand{\set}[1]{\left\{#1\right\}}

\newcommand{\Maeig}{\ensuremath{\lambda_{\mbox{\small max}}}}
\newcommand{\transpose}{\rm T}
\newcommand{\jw}{\ensuremath{\jmath\omega}}
\DeclareMathOperator{\rank}{rank}         

\usepackage{psfrag,color}
\usepackage{enumerate,cite,latexsym,graphicx}
\newtheorem{theorem}{Theorem}
\newtheorem{lemma}{Lemma}

\newtheorem{definition}{Definition}

\newtheorem{assumption}{Assumption}
\newtheorem{remark}{Remark}

\title{\LARGE \bf Negative Imaginary Systems Theory in the Robust Control of Highly Resonant Flexible Structures
}

\author{Ian R.~Petersen%
\thanks{Research supported by the
Australian Research Council (ARC) 
}%
\thanks{Ian R. Petersen is with the School of Engineering and Information Technology, 
        University of New South Wales at the Australian Defence Force Academy, Canberra ACT 2600, Australia.
         {\tt\small i.r.petersen@gmail.com} } }%

\begin{document}

\maketitle
\thispagestyle{empty}
\pagestyle{empty}

\begin{abstract}
This paper covers recent developments in the theory of negative imaginary systems and their application to the control of highly resonant flexible structures. The theory of negative imaginary systems arose out of a desire to unify a number of classical methods for the control of lightly damped structures with collocated force actuators and position sensors including positive position feedback and integral force feedback. The key result is a stability result which shows why these methods are guaranteed to yield robust closed loop stability in the face of unmodelled spillover dynamics. Related results to be presented  connect the theory of negative imaginary systems to positive real systems theory and a negative imaginary lemma has been established which is analogous to the positive real lemma. The paper also presents recent controller synthesis results based on the theory of negative imaginary systems.
\end{abstract}

\section{Introduction} \label{sec:intro}

The problem of vibration control for flexible structures arises in a variety of  applications such as large space structures,  flexible dynamics in air vehicles, nano-positioning, the control of atomic force microscopes and the control of optical interfermetric sensing systems; e.g., see \cite{PRE02,DEM07}.
A particular problem in vibration control for flexible structures is that unmodelled  (spillover) dynamics can severely degrade the control system performance or lead to instability.  
Also, uncertainties in mode frequencies and damping can lead to similar problems of poor control system performance or instability; e.g., see  \cite{BM90,DLSS08}.

An  approach to the control of flexible structures which addresses these robustness issues is {\it
  positive-position feedback}, which involves the use of collocated
force actuators and position sensors \cite{GC85,FC90}. Roughly speaking a positive position feedback controller consists of a transfer
function matrix which has the same form as a finite mode model of a
flexible structure. 
This approach was originally proposed as an alternative to passivity
based velocity feedback approaches in the control of large space
structures. 
The original papers on positive position feedback included stability
proofs based on Routh-Hurwitz and Nyquist techniques but the robustness
properties were never fully quantified. 
The research presented in this paper aims to develop a general systems theory framework for
positive position control and related control methods for flexible
structures. 
An important special case of positive-position feedback involves the
use of piezoelectric actuators and sensors. 
A positive-position feedback controller can be designed to increase the damping of the modes of a flexible structure.
Furthermore, this controller is robust against
uncertainty in the modal frequencies as well as
unmodelled plant dynamics.
We consider the theory of negative-imaginary
systems  to reveal the robustness properties of MIMO positive-position feedback controllers and related types of controllers for flexible
structures; e.g., see \cite{HM01,PMS02,PRE02,AFM07a,BMP1a,MM05,MVB06,CH10a,MMB11,LP_CSM}.

The negative-imaginary property of linear systems can be extended to nonlinear systems through the notion of  counterclockwise input-output dynamics
\cite{ANG04,ANG06,CH10}. It is shown in \cite{POB05} for
the single-input, single-output (SISO) linear case  that  the results of \cite{ANG04,ANG06} guarantee the stability  of a
positive-position feedback control system in the presence of  unmodelled dynamics and parameter
uncertainties that maintain the negative-imaginary property of the plant.

\subsection{Flexible Structure Modeling}
In modeling an undamped flexible structure with  collocated force
actuators and position sensors,
 modal analysis
 can be applied to  the relevant partial differential equation
 leading to the
transfer function matrix
\[
P(s) = \sum_{i=1}^\infty \frac{1}{s^2 + \kappa_i s
  +\omega_i^2}\psi_i \psi_i^{\transpose},
\]
where, for all $i$, $\kappa_i > 0$, $\omega_i > 0$, and $\psi_i$ is an
$m \times 1$ vector.
Therefore, the {\em Hermitian-imaginary part} \[\Im_{\text{H}}[P(\jw)] = -\frac{1}{2}\jmath(P(\jw) - P^{*}(\jw))\] of the frequency response matrix
$P(\jw)$
satisfies
\[
\Im_{\text{H}}[P(\jw)]
= -\omega\sum_{i=1}^\infty \frac{
  \kappa_i}{\left(\omega_i^2-\omega^2\right)^2 + \omega^2
  \kappa_i^2}\psi_i \psi_i^{\transpose}
\leq 0
\]
for all $\omega\geq 0.$
That is, the frequency response
 matrix 
 has a negative-semidefinite Hermitian-imaginary part for all
$\omega \geq  0$.  
We thus refer to the
transfer function matrix $P(s)$ 
as negative imaginary. 
Any flexible structure with collocated force actuators and position
sensors will have a negative imaginary transfer function matrix.

\section{Negative Imaginary Transfer Function Matrices}
\label{sec:problem-formulation}
In this section, we present the formal definition of a negative imaginary transfer function matrix. We also present a number of extensions to this notion, which have appeared in the literature. 

\begin{figure}[htbp]
\psfrag{M(jw)}{$R(\jw)$}
\centering
\includegraphics[width=8cm]{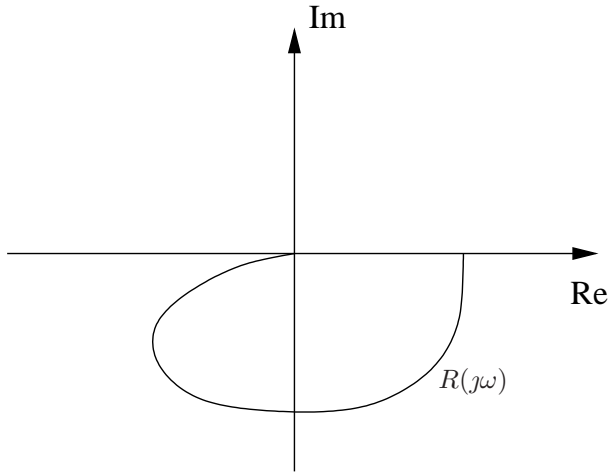}
\caption{Nyquist plot of a negative-imaginary system.}
\label{one-NIFR-system}
\end{figure}

\begin{definition}[\cite{LP06a,LP06,XiPL1a,XiPL1}]
  \label{dfn:1}
  A square real-rational proper transfer function matrix $R(s)$ is
  termed \emph{negative imaginary} (NI) if
  \begin{enumerate}
  \item $R(s)$ has no poles at the origin and in $\Re[s]>0$;
  \item $j[R(j\omega)-R^*(j\omega)]\ge0$ for all $\omega\in(0,\infty)$ except
    values of $\omega$ where $j\omega$ is a pole of $R(s)$;
  \item If $j\omega_0$, $\omega_{0}\in(0,\infty)$, is a pole of $R(s)$, it is
    at most a simple pole, and the residue matrix $K_0\triangleq\lim_{s\to
      j\omega_0}(s-j\omega_0)jR(s)$ is positive semidefinite Hermitian.
  \end{enumerate}
\end{definition}
In the SISO case, a transfer function is negative imaginary if and only if it
has no poles in open right half plane (ORHP) or at the origin and its
phase  is
in the interval $[-\pi,~0]$ at all frequencies. Consequently, the  Nyquist
plot of a SISO negative-imaginary transfer function  lies below
the real axis; see Figure \ref{one-NIFR-system}. 

\begin{remark}
Note that this definition has been recently extended to allow for a simple pole at the origin; \cite{MKPL2a,MKPL4a}. However, in this case, this leads to a slightly different form of the stability condition. Also, in \cite{XiPL2a}, a lossless version of this definition is given along with a corresponding negative imaginary lemma. 
\end{remark}

The following lemma from \cite{LP06,XiPL1} describes the  relationship between negative
imaginary transfer function matrices and positive real transfer function
matrices.

\begin{lemma}[\cite{LP06,XiPL1}]
  \label{lem:3}
  Given a square real-rational strictly proper transfer function matrix
  $R(s)$. Then $R(s)$ is negative imaginary if and only if
  \begin{enumerate}
  \item $R(s)$ has no poles at the origin;
  \item $F(s)\triangleq sR(s)$ is positive real.
  \end{enumerate}
\end{lemma}

The following definition defines the strict version of the negative imaginary property. 

\begin{definition}[See \cite{LP06a,LP06}.]
  \label{dfn:2}
  A square real-rational proper transfer function matrix $R(s)$ is
  termed \emph{strictly negative imaginary} (SNI) if
  \begin{enumerate}
  \item $R(s)$ has no poles in $\Re[s]\ge0$;
  \item $j[R(j\omega)-R^*(j\omega)]>0$ for $\omega\in(0,\infty)$.
  \end{enumerate}
\end{definition}

\begin{remark}
In the paper \cite{LSPP1}, the strict negative imaginary property is extended to a notion of strongly strict negative imaginary (SSNI) systems by imposing the additional conditions that 
\[
\displaystyle{\lim_{\omega\rightarrow \infty}}j\omega(R(j\omega)-R(j\omega)^{*})>0
\]
and
\[
\displaystyle{\lim_{\omega\rightarrow 0}}j\frac{1}{\omega}(R(j\omega)-R(j\omega)^{*})>0.
\]
This extension is useful in characterizing the strict negative imaginary property using LMIs and in controller synthesis. 
\end{remark}

The following lemmas state some useful properties of negative imaginary transfer
function matrices.

\begin{lemma}[\cite{LP06,XiPL1}]
  \label{lem:4}
  A square real-rational proper transfer function matrix $R(s)$ is
  negative imaginary if and only if $R(\infty)=R^{T}(\infty)$ and
  $\hat{R}(s)\triangleq R(s)-R(\infty)$ is negative imaginary.
\end{lemma}

\begin{lemma}[\cite{LP_CSM}]
\label{L1}
If the
$m\times m$ transfer function matrix $R(s)$
is   NI and the
$m\times m$ transfer function matrix $R_s(s)$ is NI, then the transfer function matrix
\begin{equation}
\label{sum_ni}
R_T(s) = R(s)+R_s(s)
\end{equation}
is   NI. Also, if in addition $R_s(s)$ is SNI, then $R_T(s)$ is SNI.
\end{lemma}

\begin{lemma}[\cite{LP_CSM}]
\label{LL2}
Consider the NI transfer function matrices
  $R(s)$ and $R_s(s)$, and suppose that the positive-feedback interconnection
of this transfer functions 
is internally stable. Then the corresponding $2m\times 2m$ closed-loop transfer function matrix
\begin{eqnarray}
\label{T}
T(s) &=& \left[\begin{array}{c} R(s)\left(I-R_s(s)R(s)\right)^{-1}  \\
R_s(s)\left(I-R(s)R_s(s)\right)^{-1}R(s) 
\end{array}\right. \nonumber \\
&& \left.\begin{array}{c}
R(s)\left(I-R_s(s)R(s)\right)^{-1}R_s(s) \\
R_s(s)\left(I-R(s)R_s(s)\right)^{-1}
\end{array}\right]
\end{eqnarray}
 is NI. Furthermore, if in addition, either
$R(s)$ or $R_s(s)$ is  SNI, then $T(s)$ is SNI.
\end{lemma}

We can also can also extend the above definitions of NI systems to the finite bandwidth case.

\begin{definition}[\cite{XiPL3a}]
  \label{dfn:3}
  A square real-rational proper transfer function matrix $R(s)$ is said to be
  finite frequency negative imaginary (FFNI) with bandwidth
  $\bar{\omega}$ if it satisfies the following conditions:
  \begin{enumerate}
  \item $R(s)$ has no poles at the origin and in the open right-half of the
    complex plane;
  \item $j[R(j\omega)-R^{*}(j\omega)]\ge0$ for all $\omega\in\Omega$, where \\  $\Omega=\set{\omega\in\mathbb{R} : \text{$0<\omega\le\bar{\omega}$,
        $j\omega$ is not a pole of $R(s)$}}$;
  \item Every pole of $R(s)$ on $j\bar{\Omega}$, if any, is simple and the
    corresponding residue matrix of $jR(s)$ is positive semidefinite
    Hermitian, where $\bar{\Omega}=\set{\omega\in\mathbb{R} :
      0<\omega\le\bar{\omega}}$;
  \item $R(\infty)=R^{T}(\infty)$.
  \end{enumerate}
\end{definition}

Based on the above definition, we have the following properties.
\begin{lemma}[\cite{XiPL3a}]
  \label{rem:1}
  If $R(s)$ is an FFNI transfer function matrix with bandwidth $\bar{\omega}$, 
  then the following properties hold:
  \begin{enumerate}
  \item $j[R(j\omega)-R^{*}(j\omega)]\le0$ for all $\omega\in\Omega_{1}$,
    where\\
 $\Omega_{1}=\set{\omega\in\mathbb{R} :
      \text{$-\bar{\omega}\le\omega<0$, $j\omega$ is not a pole of
        $R(s)$}}$. 
  \item Every pole of $R(s)$ in $j\bar{\Omega}_{1}$, if any, is simple and the
    corresponding residue matrix of $jR(s)$ is negative semidefinite
    Hermitian, where $\bar{\Omega}_{1}=\set{\omega\in\mathbb{R} :
      -\bar{\omega}\le\omega<0}$. 
  \end{enumerate}
\end{lemma}

Note that the stability result given in \cite{PL11}, uses a notion similar to the FFNI property defined above along with a small gain property to establish robust stability. 

\section{Some Negative Imaginary Lemmas}

The following Theorem provides a state space condition for a system to be negative imaginary. 

\begin{theorem}[Negative Imaginary Lemma,\cite{LP06a,LP06,XiPL1a,XiPL1}.]
  \label{lem:7}
  Let $(A,B,C,D)$ be a minimal state-space realization of an $m \times m$
  real-rational proper transfer function matrix $R(s)$, where
  $A\in\mathbb{R}^{n \times n}$, $B\in\mathbb{R}^{n \times m}$,
  $C\in\mathbb{R}^{m \times n}$, $D\in\mathbb{R}^{m \times m}$. Then $R(s)$ is
  negative imaginary if and only if
  \begin{enumerate}
  \item $\det(A)\ne0$, $D=D^T$, and
  \item there exists a matrix $Y=Y^T>0$, $Y\in\mathbb{R}^{n \times n}$, such
    that
    \begin{align*}
      AY + YA^T \le 0, \quad \text{and} \quad B + AYC^T=0.
    \end{align*}
  \end{enumerate}
\end{theorem}

The following alternative form of the negative imaginary lemma also holds in the case that the system has a simple pole at the origin.

\begin{theorem}[Modified NI Lemma,\cite{XiPL1a,XiPL1,MKPL2a,MKPL4a}.]
  \label{thm:0}Let $(A,B,C,D)$ be a minimal realization of $R(s)\in\mathcal{R}^{m\times m}$, where $A\in\mathbb{R}^{n\times n}$, $B\in\mathbb{R}^{n\times m}$, $C\in\mathbb{R}^{m\times n}$, $D\in\mathbb{R}^{m\times m}$, $m\leq n$.
Then $R(s)$ is negative-imaginary if and only if  $D=D^{T}$ and there exist matrices 
$P=P^T \geq 0$, $P\in\mathbb{R}^{n \times n}$, $W \in \mathbb{R}^{m\times m}$, and $L \in  \mathbb{R}^{m\times n}$ such that the following LMI is satisfied:
\begin{eqnarray}\label{NILlmi}
    \lefteqn{\left[\begin{array}{cc}
            PA+A^{T}P & PB-A^{T}C^{T} \\
           B^{T}P-CA & -(CB+B^{T}C^{T})
          \end{array}
    \right]} \nonumber \\
&& = \left[\begin{array}{cc}
-L^TL & -L^T W \\
-W^TL & -W^TW
 \end{array}
    \right]
\leq 0.
\end{eqnarray}
\end{theorem}

\begin{remark}
Note that a version of this negative imaginary lemma for descriptor systems is presented in \cite{MKPL5a}. 
\end{remark}

The following theorems provide corresponding conditions for strict negative imaginary properties.

\begin{theorem}[Weak SNI Lemma,\cite{XiPL1a,XiPL1}.]
  \label{lem:8}
  Let $(A,B,C,D)$ be a minimal state-space realization of an $m \times m$
  real-rational proper transfer function matrix $R(s)$, where
  $A\in\mathbb{R}^{n \times n}$, $B\in\mathbb{R}^{n \times m}$,
  $C\in\mathbb{R}^{m \times n}$, $D\in\mathbb{R}^{m \times m}$. Then $R(s)$ is
  strictly negative imaginary if and only if
  \begin{enumerate}
  \item $A$ is Hurwitz, $D=D^T$, $\rank(B)=\rank(C)=m$;
  \item There exists a matrix $Y=Y^T>0$, $Y\in\mathbb{R}^{n \times n}$, such
    that
    \begin{align*}
      AY + YA^T \le 0, \quad \text{and} \quad B + AYC^T=0;
    \end{align*}
  \item The transfer function matrix $M(s)\sim \left[
      \begin{array}{c|c}
        A & B \\ \hline LY^{-1}A^{-1} & 0
      \end{array}
    \right]$ has full column rank at $s=j\omega$ for any
    $\omega\in(0,\infty)$. Here
    $L^TL=-AY-YA^T$. That is, $\rank(M(j\omega))=m$ for any $\omega\in(0,\infty)$.
  \end{enumerate}
\end{theorem}

The following theorem provides a purely LMI approach to test the strong SNI property. 

\begin{theorem}\label{SSNI Lemma}[Strong SNI Lemma, \cite{LSPP1, SPLP1a}.]
Given a square transfer function matrix $R(s)\in\mathcal{R}^{m\times m}$ with a state-space realization $(A,B,C,D)$, where $A\in\mathbb{R}^{n\times n}$, $B\in\mathbb{R}^{n\times m}$, $C\in\mathbb{R}^{m\times n}$ and $D\in\mathbb{R}^{m\times m}$. Suppose $R(s)+R(-s)^{T}$ has normal rank $m$ and $(C,A)$ is observable. Then, $A$ is Hurwitz and $R(s)$ is Strongly Strict Negative Imaginary if and only if $D=D^{T}$ and there exists a matrix $Y=Y^{T}>0$ such that
\begin{equation}\label{KYPSNI}
    AY+YA^{T}<0~\text{and}~ B=-AYC^{T}.
\end{equation}
\end{theorem}

The following theorem from \cite{XiPL3a} gives a finite bandwidth version of the Negative Imaginary Lemma. 

\begin{theorem}[FFNI Lemma \cite{XiPL3a}.]
  \label{thm:1}
  Consider a real-rational proper \linebreak transfer function matrix $R(s)$ with a
  minimal state-space realization $(A,B,C,D)$. Suppose all poles of $R(s)$ are
  in the closed left-half of the complex plane, and the poles on the imaginary
  axis, if any, are simple. Let a positive scalar $\bar{\omega}$ be given. Also, suppose that if
  $A$ has eigenvalues $j\omega_{i}$ ($i\in\set{1,\ldots,q}$) such that
  $0<\omega_{i}\le\bar{\omega}$, the residue of $(sI-A)^{-1}$ at $s=jw_{i}$ is
  given by $\Phi_{i}\triangleq \lim_{s\to j\omega_{i}}
  (s-j\omega_{i})(sI-A)^{-1}$.  Then the following statements are equivalent:
  \begin{enumerate}
  \item The transfer function matrix $R(s)$ is FFNI with bandwidth $\bar{\omega}$.
  \item $\det(A)\ne0$, $D=D^{T}$, and the transfer function matrix $F(s)$ with
    a minimal state-space realization $(A,B,CA,CB)$ is FFPR with bandwidth
    $\bar{\omega}$.
  \item $\det(A)\ne0$, $D=D^{T}$, and $CA\Phi_{i}B=(CA\Phi_{i}B)^{*}\ge0$ for all
    $i\in\set{1,\ldots,q}$ if $A$ has any eigenvalues on $j\bar{\Omega}$ where
    $\bar{\Omega}=\set{\omega\in\mathbb{R} : 0<\omega\le\bar{\omega}}$. Also, 
    there exist real symmetric matrices $P=P^{T}$ and $Q=Q^{T}\ge0$ such that
    \begin{align}
      \label{eq:1}
      PA + A^{T}P - A^{T}QA + \bar{\omega}^{2}Q &\le 0, \\
      \label{eq:2}
      C + B^{T}A^{-T}P &= 0, \\
      \label{eq:3}
      QA^{-1}B &= 0.
    \end{align}
  \item $\det(A)\ne0$, $D=D^{T}$, and  $CA\Phi_{i}B=(CA\Phi_{i}B)^{*}\ge0$ for all
    $i\in\set{1,\ldots,q}$ if $A$ has any eigenvalues on $j\bar{\Omega}$ where
    $\bar{\Omega}=\set{\omega\in\mathbb{R} : 0<\omega\le\bar{\omega}}$. Also,
    there exist real symmetric matrices $Y=Y^{T}$ and $X=X^{T}\ge0$ such that
    \begin{align}
      \label{eq:4}
      AY + YA^{T} - AXA^{T} + \bar{\omega}^{2}X &\le 0, \\
      \label{eq:5}
      B + AYC^{T} &= 0, \\
      \label{eq:6}
      CX &= 0.
    \end{align}
  \end{enumerate}
\end{theorem}

In addition to the above negative imaginary lemmas, which involve LMI conditions to test if a given system is negative imaginary or strictly negative imaginary, the following result uses an eigenvalue test of a corresponding Hamiltonian matrix to test if a given system is NI. 

\begin{theorem}[\cite{MKPL1a}]
\label{th:ni-thm} Consider a transfer function matrix $R(s)$ with a minimal realization $(A,B,C,D)$.  Also, suppose $A$ is a
Hurwitz matrix, $Q=CB+B^{T}C^{T}> 0$, and $D=D^{T}$. Then, $R(s)$ is NI if
and only if the following conditions are satisfied:
\begin{enumerate}
\item[1.] The Hamiltonian matrix,
\begin{equation}
\label{eq:hamiltonian-mtx}
    N=
    \begin{bmatrix}
        -A+BQ^{-1}CA & BQ^{-1}B^{T} \\
        -A^{T}C^{T}Q^{-1}CA & A^{T}-A^{T}C^{T}Q^{-1}B^{T}
    \end{bmatrix} \nonumber
\end{equation}
 has no pure imaginary eigenvalues with odd multiplicity.

\item[2.] $j\frac{\left(
\omega_{i}+\omega _{i+1}\right) }{2}\left( R(j\eta _{i})-R(j\eta _{i})^{\ast }\right) $ has
only positive real eigenvalues for all $i\in \{1,k-1\}$, where
$\omega_{i}\in \Omega $, and $\Omega =\left\{ \omega _{1}, \omega
_{2}, \ldots, \omega _{k}\right\} $ is the set of frequencies listed
in strictly increasing order, such that $\det \left[ F(j\omega_{i}
)+F(j\omega_{i} )^{\ast }\right] = 0.$ Here, $F(s)= s\tilde{R}(s)$ where 
$\tilde{R}(s)=R(s)-D$.
\end{enumerate}
\end{theorem}

\begin{remark}
This approach has been extended in the paper \cite{MKPL3a} to consider the problem of perturbing a non-NI system model in order to obtain an NI system model. 
\end{remark}

Also, the paper \cite{MKPL1a} gives the following result for testing the NI property of SISO systems. 

\begin{assumption}
\label{assm:siso}
 Let $A \in \mathbb{R}^{n
\times n},B \in \mathbb{R}^{n \times 1},C \in \mathbb{R}^{1 \times
n},D \in \mathbb{R}$, and suppose the transfer function
$G(s)=C(sI-A)^{-1}B+D$ has all of its poles and zeros in the closed
left half of the complex plane excluding poles at the origin. Also,
any pole on the imaginary axis is assumed to be simple.
\end{assumption}

\begin{theorem}[\cite{MKPL1a}]
\label{thm:siso-ni} The SISO transfer function  $G(s) = C (sI -
A)^{-1} B + D$ with $CB > 0$, and satisfying  Assumption
\ref{assm:siso} is NI if and only if the following conditions are
satisfied:
    \begin{enumerate}
        \item[1.] The matrix $(I-\frac{1}{CB}BC)A^{2}$ has no eigenvalues of odd (algebraic)
multiplicity on the open negative real axis.
        \item[2.] All residues of $F(s)=s\left( G(s)-D\right) $ corresponding to  poles on the
imaginary axis are positive.
    \end{enumerate}
\end{theorem}

\section{Stability of Interconnections of Negative Imaginary Systems}
\label{sec:interc-negat-imag}

\begin{figure}[htbp]
\centering
\psfrag{M}{$R(s)$}
\psfrag{N}{$R_s(s)$}
\psfrag{w1}{$w_1$}
\psfrag{u1}{$u_1$}
\psfrag{y1}{$y_1$}
\psfrag{w2}{$w_2$}
\psfrag{u2}{$u_2$}
\psfrag{y2}{$y_2$}
\includegraphics[width=8cm]{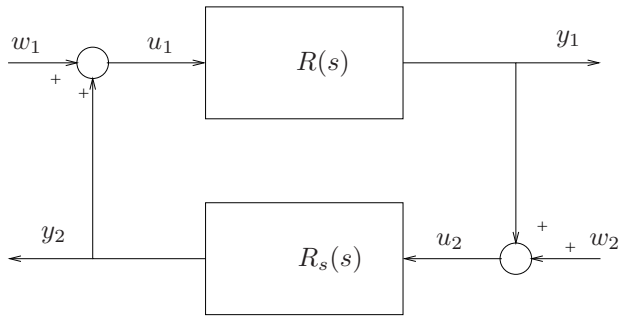}
\caption{A positive feedback interconnection.}
\label{feedback-interconnection}
\end{figure}

The following theorem, which is the main stability result in the theory of negative imaginary systems, gives a necessary and sufficient condition for the stability of the positive feedback interconnection (see Figure \ref{feedback-interconnection}) between an NI system and an SNI system. 

\begin{theorem}[\cite{LP06a,LP06,XiPL1a,XiPL1}]
  \label{thm:2}
  Given a negative imaginary transfer function matrix $R(s)$ and a strictly
  negative imaginary transfer function matrix $R_s(s)$ that also satisfy
  $R(\infty)R_s(\infty)=0$ and $R_s(\infty)\ge0$. Then the positive feedback
  interconnection $[R(s),R_s(s)]$ is internally stable if and only if
  $\lambda_{\mathrm{max}}(R(0)R_s(0)) < 1$.
\end{theorem}

\begin{remark}
Note that in the papers \cite{ELP1a,EPLP1}, the robustness properties obtained by this theorem are compared with the robustness properties which are obtained from the $\mu$ analysis approach. Also, in the papers \cite{SLPP2a,SLPP2}, this stability robustness result is extended to allow for additional norm bounded uncertain parameters. 
\end{remark}

The above theorem can also be extended to give a sufficient condition for closed loop stability in  the case in which the NI system has a simple pole at the origin.

\begin{theorem}[\cite{MKPL2a,MKPL4a}]
  \label{thm:3}
Consider  a strictly proper negative imaginary transfer function matrix $R(s)$ with minimal state space realization $(A,B,C,0)$, which may contain a simple pole at the origin. Also, consider a strictly
  negative imaginary transfer function matrix $R_s(s)$.
 Then the closed-loop positive feedback
interconnection between $R(s)$ and $R_s(s)$ is internally stable
if $R_s(s) < 0$ and the matrix $A+BR_s(0)C$ is not singular.
\end{theorem}

\section{Negative-Imaginary  Feedback  Controllers}

In this section, we present a number of different classes of commonly used negative imaginary controllers; see also \cite{LP_CSM}.

\subsection{Positive-Position Feedback Controllers}
In the SISO case, a positive-position feedback
controller (e.g., see \cite{GC85,FC90,MVB05,RBFM05,MVB06,MA09,MM09a,MM09,YAM10,AP11}) is a controller of the form
\begin{equation}
\label{siso-ppf}
C(s) = \sum_{i=1}^M\frac{k_i}{s^2 + 2\zeta_i\omega_{i} s +
  \omega_{i}^2},
\end{equation}
where $\omega_{i}>0$, $\zeta_i>0$, and $k_i>0$ for $i = 1,2,\ldots,M$. Such controllers are always SNI. 

\subsection{Resonant Controllers}
Resonant controllers are proper but not strictly proper  SISO  SNI controllers of the form
\begin{equation}
\label{resonant1}
C(s) = \sum_{i=1}^M\frac{-k_is^2}{s^2 + 2\zeta_i\omega_{i} s +
  \omega_{i}^2},
\end{equation}
 where $\omega_{i} >0$, $\zeta_i>0$, and $k_i>0$ for $i =
 1,2,\ldots,M$; e.g, see \cite{PMS02,MV04}. The controller (\ref{resonant1}) can be implemented as the
 positive-position feedback controller (\ref{siso-ppf}) using an acceleration
 sensor rather than a position sensor.  Alternatively, (\ref{resonant1}) can be
 implemented as the  positive-real feedback controller
\[
\bar C(s) = \sum_{i=1}^M\frac{k_is}{s^2 + 2\zeta_i\omega_{i} s +
  \omega_{i}^2},
\]
where $\omega_{i} >0$, $\zeta_i>0$, and $k_i>0$ for $i =
 1,2,\ldots,M$, using a velocity sensor rather than a position sensor. Such controllers are always SNI. 

\subsection{Integral Resonant Controllers}
MIMO transfer function matrices of the form
\[
C(s) = [sI+\Gamma \Phi]^{-1}\Gamma
\]
are SNI.  Here,
$\Gamma$ is a positive-definite matrix and $\Phi$ is a
positive-definite matrix. The use of a controller of this
form when applied to a flexible structure with force actuators and
position sensors is referred to as {\em integral resonant control}, or
integral force control
\cite{PRE02,AFM07a,AFM07b,PMA08,BM08,BM09,BMP1a,BMP1,PAFM11}.

\section{State-Feedback Controller Synthesis}

\begin{figure}[htbp]
\centering
\psfrag{D(s)}{$\Delta(s)$}
\psfrag{G(s)}{$G_{\mbox{\small cl}}(s)$}
\includegraphics[width=8cm]{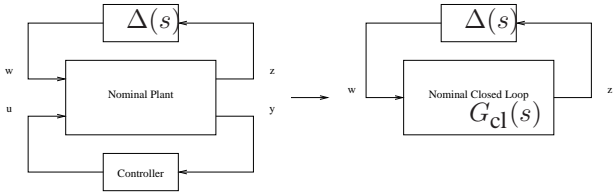}
\caption{A feedback control system.
 }
\label{F5.3}
\end{figure}

Consider the  feedback control system  in Figure
\ref{F5.3} in the case that  full state feedback is available. In
  this case, the negative imaginary lemma  can be used to synthesize a state-feedback control law such
that the resulting closed-loop system is NI. Indeed, suppose  the  uncertain system shown in Figure
\ref{F5.3} is described by the state
equations
\begin{eqnarray}
\label{sys_ol}
&\dot x = Ax +B_1w+B_2u,& \\
\label{sys_ola}
&z = C_1x,&  \\
\label{sys_olb}
&w = \Delta(s)z,&
\end{eqnarray}
where  the uncertainty transfer function matrix  $\Delta(s)$ is assumed
to be  SNI with $|\Maeig(\Delta(0))|
\leq 1$ and $\Delta(\infty) \geq 0$.
Applying the state-feedback control law $u = Kx$ yields
the closed-loop uncertain system
\begin{eqnarray}
\label{sys_cl}
&\dot x = (A+B_2K)x +B_1w,& \\
\label{sys_cla}
&z = C_1x,&  \\
\label{sys_clb}
&w = \Delta(s)z.&
\end{eqnarray}
The corresponding nominal closed-loop transfer
function matrix is
\begin{equation}
\label{Gcl}
G_{\mbox{\small cl}}(s) = C_1(sI-A-B_2K)^{-1}B_1.
\end{equation}

\begin{theorem}[\cite{PLS1a,LP_CSM}]
\label{SFNI_DCgain}
Consider the uncertain system (\ref{sys_ol}),  (\ref{sys_ola}),  (\ref{sys_olb}) and
suppose there exist  matrices $Y > 0$,
 $M$, and
 a scalar  $\varepsilon > 0$ such that
\begin{eqnarray}
\lefteqn{\left[\begin{array}{c}
\label{lmi2}
AY+YA^{\transpose}+ B_2M+M^{\transpose}B_2^{\transpose} + \varepsilon I  \\
B_1^{\transpose}+C_1YA^{\transpose}+C_1M^{\transpose}B_2^{\transpose} 
\end{array}\right.} \nonumber \\
&&\left.\begin{array}{c}
 B_1+AYC_1^{\transpose}+B_2MC_1^{\transpose} \\
  0
\end{array}\right] \nonumber \\
&& \leq   0,\\
\label{lmi3}
 & C_1 Y C_1^T -I <  0,& \\
\label{lmi4}
&Y > 0.&
\end{eqnarray}
Here the parameter $\varepsilon > 0$ is chosen to be sufficiently small.
Then the state-feedback control law $u =  MY^{-1}x$ is robustly
stabilizing for the uncertain system (\ref{sys_ol}),  (\ref{sys_ola}),  (\ref{sys_olb}).
\end{theorem}

\begin{remark}
Note that in the papers \cite{LSP1a,SLPP1}, an alternative approach to controller synthesis is considered by connecting the negative imaginary synthesis problem to the $H^\infty$ control problem. 
\end{remark}

\section{Conclusions}
We have surveyed some recent results in the area of 
negative imaginary systems theory. This results characterize the class of negative imaginary systems and give methods for testing if a given system is negative imaginary. 
These results also show that the
class of NI systems  yields a
robust stability analysis result, which broadly speaking can be captured
by saying that if one system is negative imaginary and the other system
is strictly negative imaginary, then a necessary and sufficient
condition for internal stability of the positive-feedback
interconnection of  the two  systems is  that  the dc loop gain
is less than unity. 
This result provides a  
framework for the analysis of robust stability of lightly damped flexible structures
with unmodelled dynamics. 


\end{document}